\documentclass{jpsj-suppl}
\usepackage{txfonts} 

\title{JINA-NuGrid Galactic Chemical Evolution Pipeline}

\author{Benoit \textsc{C\^ot\'e}$^{1,2,8,9}$, Christian Ritter$^{1,8,9}$, Falk Herwig$^{1,8,9}$, Brian W. O'Shea$^{3,4,8}$, Marco Pignatari$^{5,9}$, Devin Silvia$^{3,10}$, Samuel Jones$^{6,9}$ and Chris L. Fryer$^{7,9}$}

\inst{$^{1}$Department of Physics and Astronomy, University of Victoria, BC, V8W 2Y2, Canada \\
$^{2}$National Superconducting Cyclotron Laboratory, Michigan State University, MI, 48823, USA \\
$^{3}$Department of Physics and Astronomy, Michigan State University, MI, 48823, USA \\
$^{4}$Department of Computational Math., Sci. and Eng., Michigan State University, MI, 48823, USA \\
$^{5}$E.A. Milne Centre for Astrophysics, Department of Phys. and Math., University of Hull, Hull, UK \\
$^{6}$Heidelberg Institute for Theoretical Studies, Heidelberg, Germany \\
$^{7}$Computational Physics and Methods, Los Alamos National Laboratory, Los Alamos, USA \\
$^{8}$JINA-CEE, Michigan State University, MI, 48823, USA, www.jinaweb.org \\
$^{9}$NuGrid Collaboration, www.nugridstars.org \\
$^{10}$National Science Foundation, Astronomy and Astrophysics Postdoctoral Fellow, USA}

\email{bcote@uvic.ca}

\recdate{August 15, 2016}

\abst{Galactic chemical evolution is a topic that involves nuclear physics, stellar evolution, galaxy evolution, observation, and cosmology.  Continuous communication and feedback between these fields is a key component in improving our understanding of how galaxies form and how elements are created and recycled in galaxies and intergalactic space.  In this proceedings, we present the current state of the JINA-NuGrid chemical evolution pipeline.  It is designed to probe the impact of nuclear astrophysics uncertainties on galactic chemical evolution, to improve our knowledges regarding the origin of the elements in a cosmological context, and to create the required interdisciplinary connections.}

\kword{Chemical evolution, stellar yields, galaxies}

\begin{document}
\maketitle

\section{Introduction}
\begin{figure*}
\begin{center}
\includegraphics[width=6.0in]{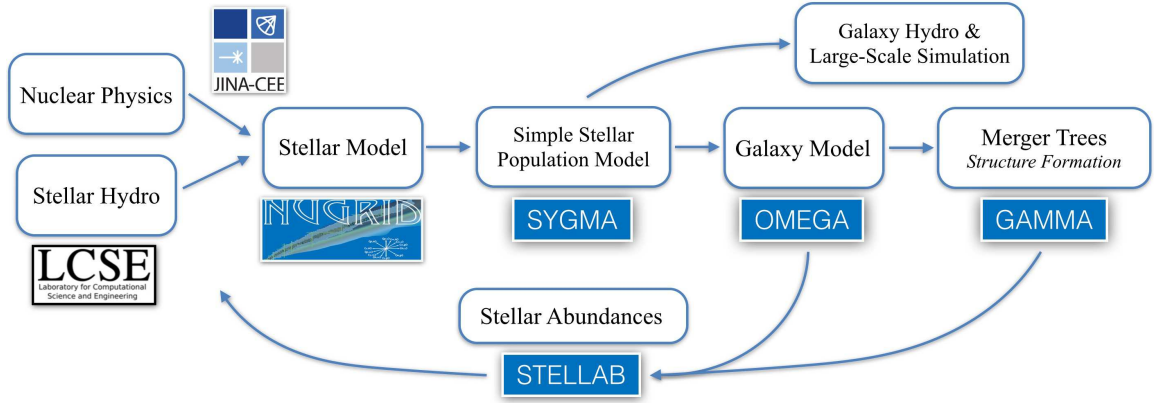}
\caption{JINA-NuGrid chemical evolution pipeline. See the text for more information on the different codes.}
\label{fig_pipeline}
\end{center}
\end{figure*}

Detailed stellar abundances from modern astronomical surveys represent a unique window on the early universe and on the formation of the Milky Way and its satellite galaxies.  However, the theoretical tools needed to translate these abundances into meaningful understanding of cosmological structure formation are challenged by uncertainties in their inputs.  In this proceedings, we present an overview of our efforts to address this fundamental challenge by creating a flexible pipeline (see Fig.~\ref{fig_pipeline}) for modeling chemical evolution and by quantifying the robustness of our chemical evolution predictions due to uncertainties in nuclear physics, stellar evolution models, and observational inputs.

\section{Input Physics}
\label{sect_input_physics}
Stellar yields are the foundation of galactic chemical evolution (GCE) models and simulations and are the result of combined efforts between nuclear and stellar physics.  Although the yields used in GCE studies are calculated with 1D stellar evolution codes, they are usually constrained by more sophisticated multi-dimensional hydrodynamic simulations.  As part of the JINA-CEE and NuGrid collaborations, the strength of our pipeline is its connection with nuclear astrophysics.  As described below, this allows us to quantify the impact of nuclear astrophysics in a GCE context.  By default, our pipeline uses the consistent set of NuGrid stellar yields \cite{p13,r16}, which provides 83 elements and 280 stables isotopes, but other yields from different research groups are also available.

\section{NuPyCEE and Statistical Tools}
\label{sect_nupycee}
The NuPyCEE (NuGrid Python Chemical Evolution Environment) package is available online at www.github.com/NuGrid/NUPYCEE and includes the three codes described below.  Userguide IPython notebooks and widgets are available at www.nugridstars.org/projects/wendi.

SYGMA \cite{r16} (Stellar Yields for Galactic Modeling Applications) folds stellar yields into simple stellar population models to provide the composition of their ejecta as a function of time and metallicity.  It includes several analysis tools and can print input tables for hydrodynamic galaxy simulations and semi-analytical models to generate stellar feedback.  All of the input parameters can be easily modified.

OMEGA \cite{c16a} (One-zone Model for the Evolution of GAlaxies) is a classical one-zone GCE code that includes several prescriptions for galactic inflows and outflows.  Using an input star formation history, it follows the contribution of several stellar populations, all created by SYGMA.  While OMEGA is not explicitly designed to provide direct insight into galaxy evolution, it is ideal to test modeling assumptions and uncertainties in stellar models \cite{c16a}.  As an example, panels \textit{a} and \textit{b} of Figure~\ref{fig_results} show the impact of using different approaches for calculating the explosive yields of massive stars.

STELLAB (Stellar Abundances) allows to compare GCE predictions with observations.  It consists of a collection of stellar abundances taken from the literature for the Milky Way and neighbouring satellite galaxies.  New observations can easily be included and subsamples can be selected to create personal dataset.  All data can be re-normalized to any solar abundances.

Our pipeline also includes statistical tools.  Our Monte Carlo code can quantify the propagation of uncertainties \cite{c16b} (panel \textit{c} of Fig.~\ref{fig_results}) and randomly sample the stellar initial mass function in order to generate scatter at low metallicity (panel \textit{d} of Fig.~\ref{fig_results}).  The coupling between OMEGA and a Markov Chain Monte Carlo (MCMC) code \cite{mcmc} can fine-tune our input parameters and provide their confidence levels for our best GCE fits (panel \textit{e} of Fig.~\ref{fig_results}), given the uncertainties in observational data \cite{c16a}.

\section{Cosmological Structure Formation}
\begin{figure*}[tbh]
\begin{center}
\includegraphics[width=6.0in]{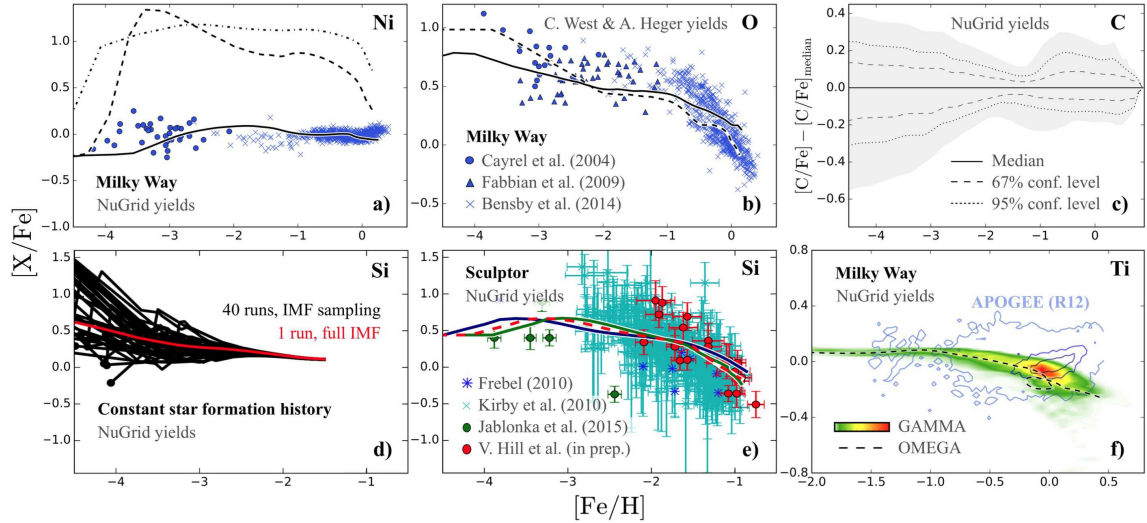}
\caption{Lines and symbols represent, respectively, GCE predictions and observational data, except for APOGEE data \cite{apogee} which are shown as a density map. \textbf{Panel a)} Impact of different mass-cuts \cite{r16}. \textbf{Panel b)} Impact of different explodability criteria \cite{c16c}. \textbf{Panel c)} Lower limit uncertainties in GCE predictions \cite{c16b}. \textbf{Panel d)} Scatter caused by randomly sampling the stellar initial mass function. \textbf{Panel e)} Best fits derived from MCMC calculations for different inflow and outflow prescriptions \cite{c16a}. \textbf{Panel f)} Prediction comparison between GAMMA and OMEGA using the Caterpillar merger-tree \#18 \cite{g16}.}
\label{fig_results}
\end{center}
\end{figure*}

To account for the hierarchical formation nature of a galaxy, GAMMA (Galaxy Assembly with Merger-trees for Modeling Abundances) uses the merger trees extracted from large-scale cosmological simulations and applies OMEGA on top each branch to follow the chemical evolution of all progenitor galaxies before they merge.  The scatter seen in GAMMA predictions is caused by the different chemical evolution histories of progenitor galaxies (panel \textit{f} of Fig.~\ref{fig_results}).  This  framework is in development and will be used to test different mass assembly histories for the Milky Way \cite{g16}, to post-process hydrodynamic simulations of the early Universe, and to study the impact of nuclear astrophysics uncertainties within a cosmological context.

\section{Future Directives}
In the near future, we plan to update our GCE pipeline by adding inhomogeneous mixing prescriptions, including different sources of heavy elements (e.g., compact binary mergers, neutrino-driven winds), and creating multi-zone models for individual galaxies and progenitors to better capture the effect of star formation on the physical state of the interstellar and intergalactic media.  Our multidisciplinary pipeline offers the opportunity to improve our knowledges regarding the origin of the elements at nuclear and stellar scales and on the relation between the chemical evolution signatures of galaxies and their mass assembly history.


\begin{thebibliography}{9}
\bibitem{p13} M. Pignatari, et al., 2013, arXiv:1307.6961
\bibitem{r16} C. Ritter, et al. 2016, in preparation
\bibitem{c16a} B. C\^ot\'e, B. W. O'Shea, C. Ritter, F. Herwig, and K. A. Venn,  2016a, arXiv:1604.07824
\bibitem{c16c} B. C\^ot\'e, C. West, A. Heger, C. Ritter, B. W. O'Shea, F. Herwig, C. Travaglio, and S. Bisterzo, 2016b, MNRAS, in press, arXiv:1602.04824
\bibitem{c16b} B. C\^ot\'e, C. Ritter, B. W. O'Shea, F. Herwig, M. Pignatari, S. Jones, and C. L. Fryer, 2016c, ApJ, 824, 82
\bibitem{mcmc} D. Foreman-Mackey, D. W. Hogg, D. Lang, and J. Goodman, 2013, PASP, 125, 306
\bibitem{g16} B. F. Griffen, A. P. Ji, G. A. Dooley, F. A. G\'omez, M. Vogelsberger, B. W. O'Shea, and A. Frebel, 2016, ApJ, 818, 10
\bibitem{apogee} APOGEE (R12) www.sdss.org/dr12/irspec/abundances
\end{thebibliography}
\end{document}